\documentclass[12pt]{article}
\usepackage{graphicx}
\begin{document} 

\centerline{\bf
Monte Carlo simulation of the rise and the fall of languages}

\bigskip

\centerline{Christian Schulze and Dietrich Stauffer}

\centerline{Institute for Theoretical Physics, Cologne University\\D-50923 K\"oln, Euroland}

\bigskip
\centerline{e-mail: stauffer@thp.uni-koeln.de}

\bigskip
Abstract: 
Similar to biological evolution and speciation we define a 
language through a string of 8 or 16 bits.  The parent 
gives its language to its children, apart from a random mutation from zero to one 
or from one to zero; initially all bits are zero. The Verhulst deaths are taken
as proportional to the total number of people, while in addition languages
spoken by many people are preferred over
small languages. For a fixed population size, a sharp phase 
transition is observed: For low mutation rates, one language 
contains nearly all people; for high mutation rates, no language
dominates and the size distribution of languages is roughly 
log-normal as for present human languages. A simple scaling 
law is valid.

\bigskip


\bigskip

\begin{figure}[hbt]
\begin{center}
\includegraphics[angle=-90,scale=0.5]{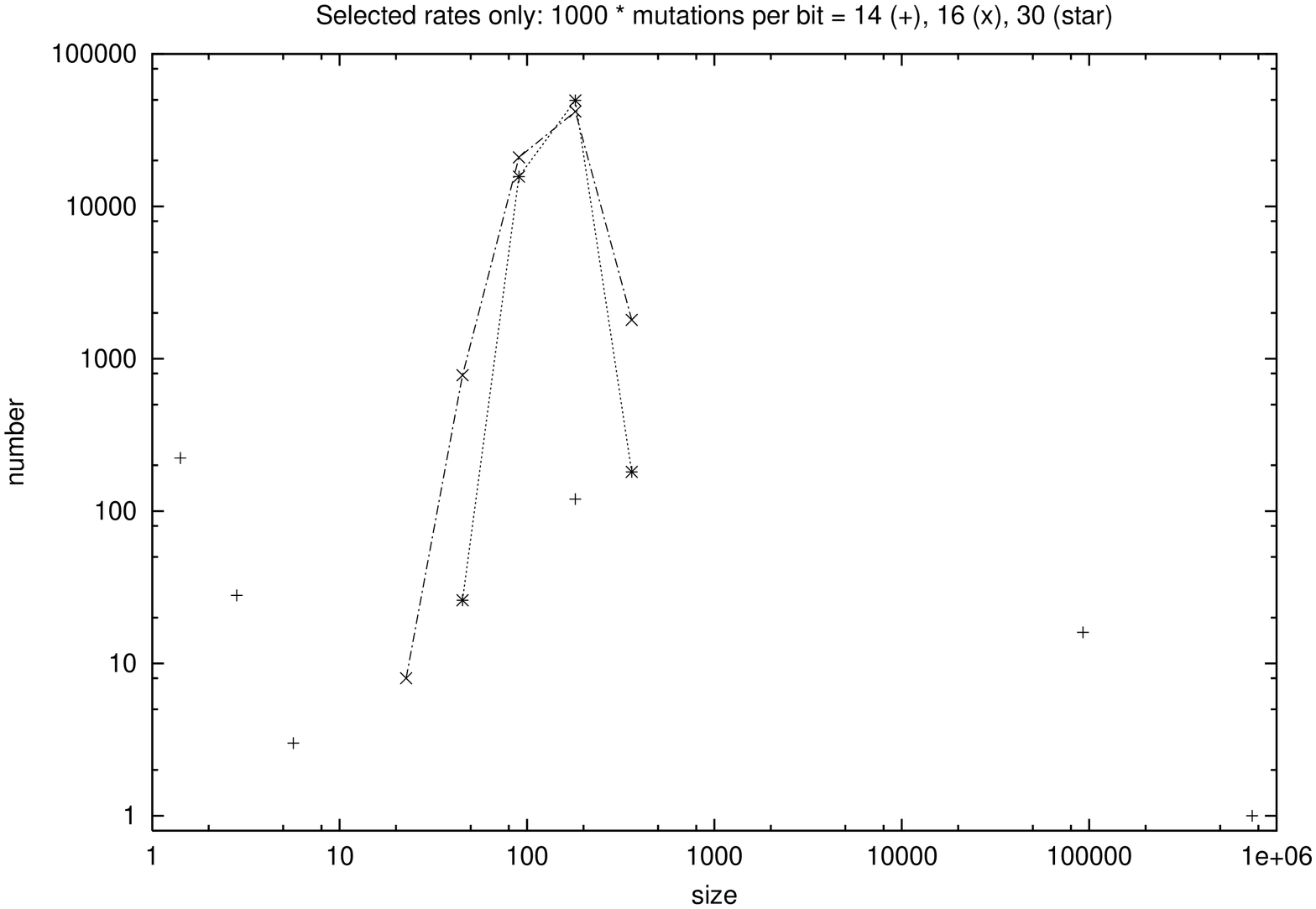}
\includegraphics[angle=-90,scale=0.5]{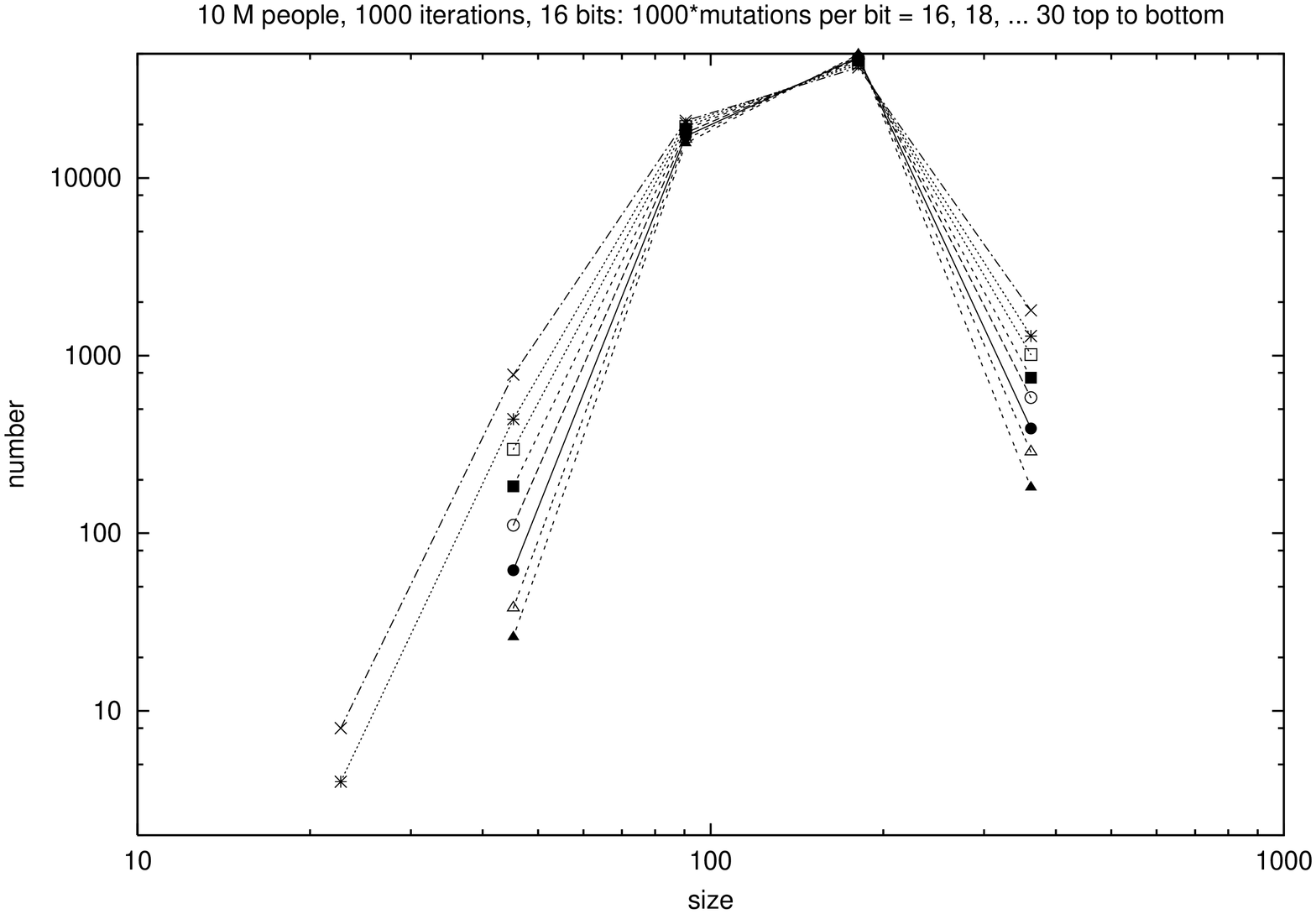}
\end{center}
\caption{ 
Histograms of language sizes for 16 bits, one sample only of
$K/2 = 10$ million people, mutations per bit 
= 0.014 (+), 0.016 ($\times$), 0.030 (stars) in part a and 0.016 to 0.030 in steps of 
0.002 in part b. 
}
\end{figure}

\begin{figure}[hbt]
\begin{center}
\includegraphics[angle=-90,scale=0.5]{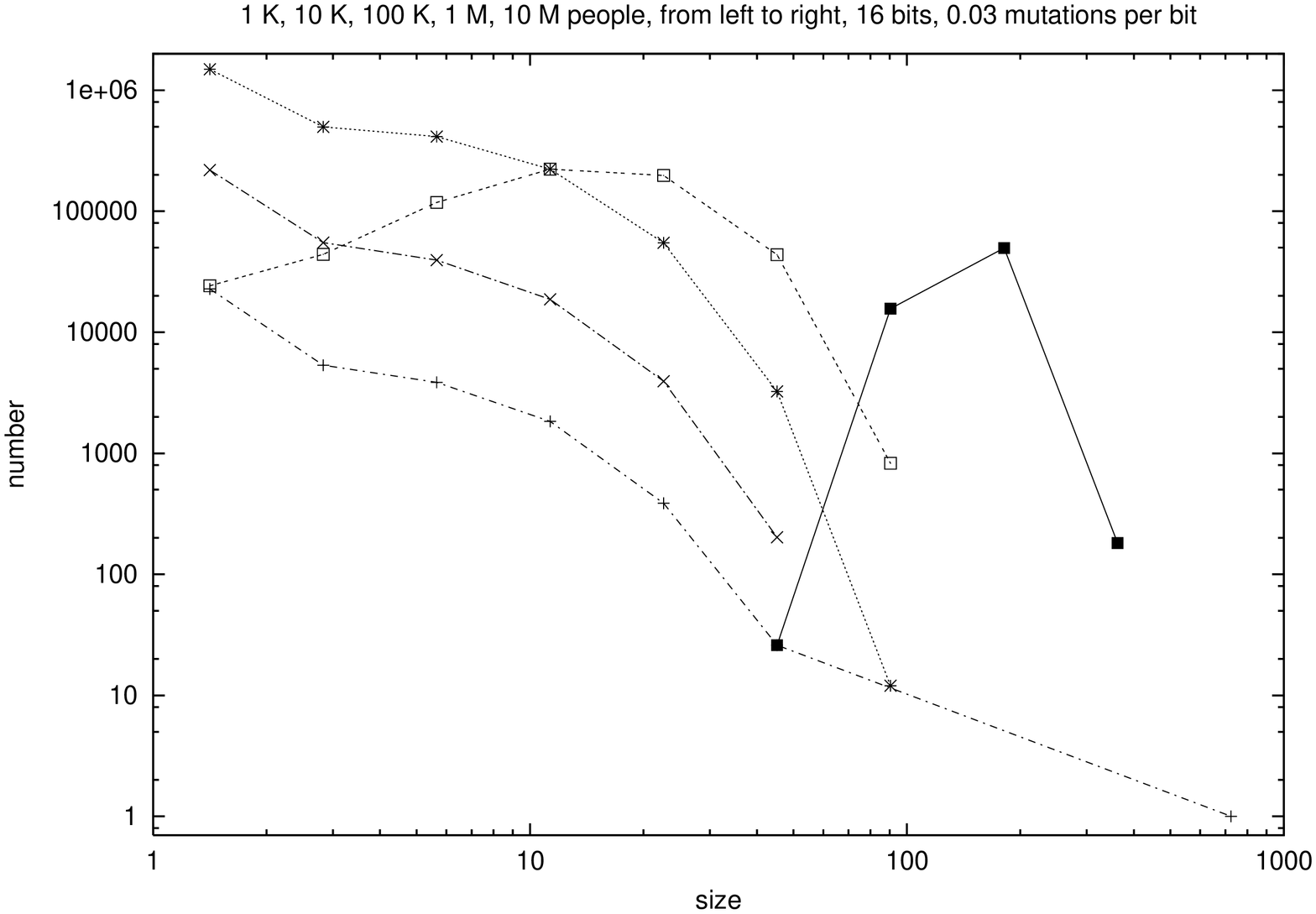}
\includegraphics[angle=-90,scale=0.5]{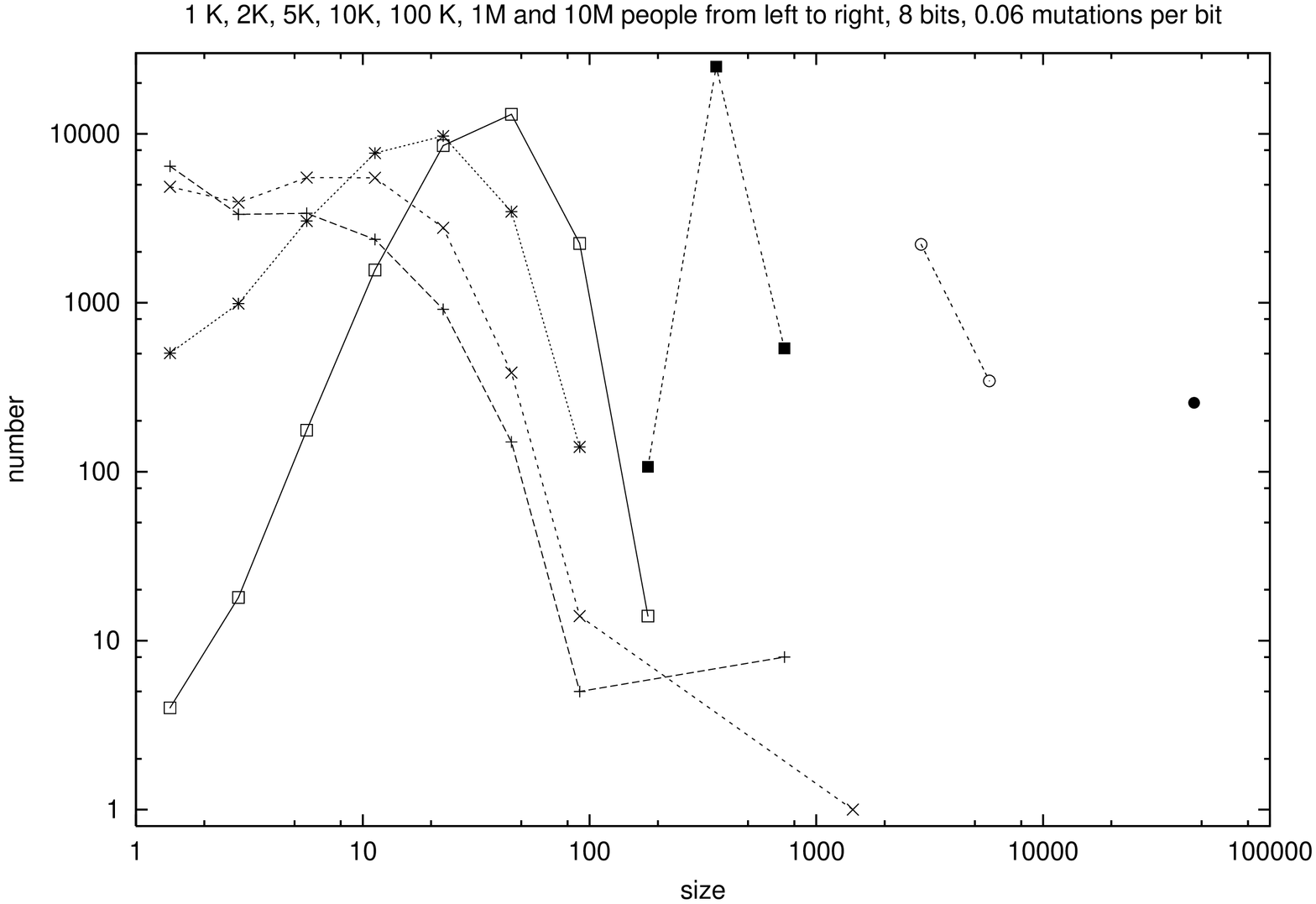}
\end{center}
\caption{ 
Histograms of languagze sizes for 16 bits (part a) and 8 bits (part b), with
same mutation rate 0.48 per individual, for different population sizes,
summed over up to 100 samples.
}
\end{figure}

\begin{figure}[hbt]
\begin{center}
\includegraphics[angle=-90,scale=0.5]{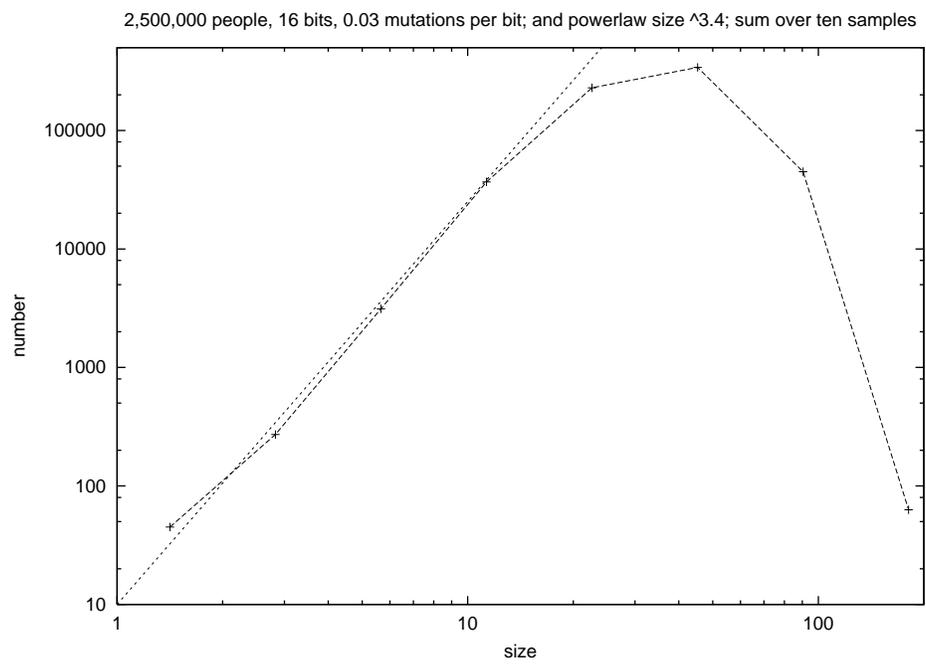}
\end{center}
\caption{ 
Roughly log-normal size distribution, with higher values for small sizes 
described by a power law.
}
\end{figure}

\begin{figure}[hbt]
\begin{center}
\includegraphics[angle=-90,scale=0.5]{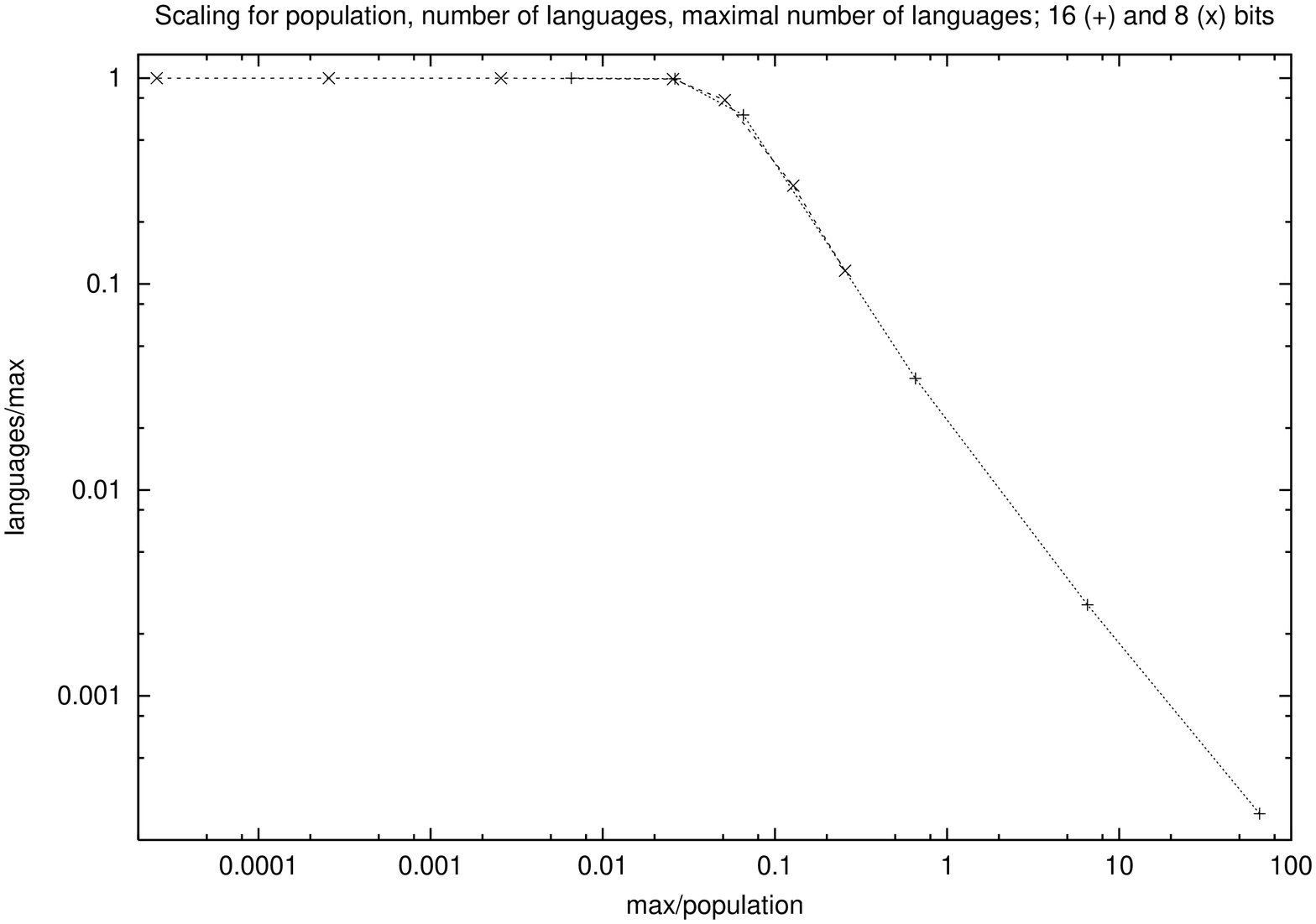}
\end{center}
\caption{ 
Scaling test: Symbols for 8 bits ($\times$) and 16 bits (+) follow the same scaling
function $f$ if plotted as $L/M$ versus $M/N_\infty$. Runs with 30 bits 
and 10 or 100 million people fit in reasonably near the lower right corner 
(not shown).
}
\end{figure}

\section{Introduction}

Human languages are grouped into families, like the Indo-European
languages, which may all have arisen from one common original 
language. For example, ancient Latin split into Portuguese, 
Spanish, French, Italian, Romanian and other languages during the
last two millenia. On the other hand, many of the present 
languages are spoken only by a relatively small number of people
and are in danger of extinction \cite{science,sutherland}.
In this way languages are similar to biological species. We
thus try to simulate languages using methods similar to the
modelling of speciation \cite{eigen,pmco}.

A language for us can be a human language (including Fortran,
...), a sign language, a system of bird songs, a human alphabet,
or any other system of communication. We simulate it by a 
string of 8, 16 or 30 bits and define languages as different if
they differ in at least one bit. The position of the bit in 
the string plays no role, in contrast to the Penna ageing model
from which program elements are taken \cite{book}.

\section{Model}

We start with one person, i.e. $N(t=0) = 1$,
speaking language zero (all bits are 
zero). Then at each iteration $t$ all $N(t)$ living people 
are subject to a Verhulst death, i.e. they die with probability 
$N(t)/K$ where $K$ in biology is often called the carrying 
capacity and incorporates the limitations of food and space. 
Each survivor produces one offspring at each iteration which
uses the same bitstring apart from one random mutation (bit 
changed from 0 to 1 or from 1 to 0) which happens with a 
probability $p$ per person (or $p/8$ per bit if the language has
8 bits). Usually, all bit-strings are assumed to be equally fit, 
in contrast to typical biological models \cite{eigen,pmco}.

Also at each iteration, each individual can switch 
from its present language to another randomly selected one,
with probability $$(2N(t)/K)(1-x^2)$$
where $x$ is the fraction of all people speaking the present 
language of that individual. The first factor, which approaches
unity for long times, ensures that at the beginning with a low
population density there is not yet much competition between 
languages, while in the later stationary high population
the less spoken languages are in danger of extinction. The 
exponent two takes into account that normally two people 
communicate with each other; thus the survival probability of
a language is proportional to the square of the number of people
speaking it.

(The final population is $K/2$ and not $K$ since we determine the Verhulst 
probability $y = N(t-1)/K$ at the beginning of iteration $t$ and leave it at
that value for the whole iteration. The Verhulst deaths thus reduce the 
population by a factor $1-y$, and if each of the survivors has $b$ offspring,
the population is multiplied by another factor $1+b$. For a stationary
population, these two factors have to cancel: $(1-y)(1+b) = 1$, giving
$y = b/(1+b) = 1/2$ for our choice $b=1$.)
 
\section{Results}

For an eventual stationary population of ten million at $t=1000$, as a 
function of increasing mutation rate $p$, a sharp transition
was observed between a dominance regime at low and a smooth 
distribution at high mutation rates $p$, Fig.1:  

i) For low $p$,
one language, usually the one with all bits zero, contains 
nearly all individuals, and the mutant languages differing 
from the dominant one by one bit only contain most of the rest.
This behaviour is hardly realistic except for alphabets. 

ii) For
high mutation rates $p$, on the other hand, no language contains
a large fraction of the population, and the distribution of 
language sizes (measured as the number of people speaking it) 
is roughly log-normal with higher statistics for small
languages. This result agrees well with reality 
\cite{sutherland}. 

In Fig.1, part a shows the drastic difference between dominance (+) and
smooth distribution ($\times$,stars), part b the slow approach to a symmetric 
log-normal distribution with increasing mutation rate. (We bin the
number of people speaking one language into powers of two, lumping 
together all languages spoken by 33 to 64 people, for example.)

In the dominance regime i) of low $p$, the number $L(t)$ of 
languages first increases from unity towards about $10^2$
and then decreases again to about a dozen (not counting languages with
less than 10 speakers). In the smooth regime
ii) of high $p$ the number $L$ of languages first increases and
then reaches a plateau, which may even equal the maximal
number $M = 2^8$ or $M = 2^{16}$ for 8 or 16 bits, respectively.

Also for a fixed mutation rate as a function of the final 
population $K/2$ we see a change from the dominance regime at 
low populations to a smooth distribution at high populations,
Fig.2. For very large populations a rather narrow distribution
of language sizes develops, i.e. the whole population is 
distributed about equally among the surviving languages. Fig.3
shows for an intermediate population a power law on the 
small-size side of the histogram, and a parabola-like curve, meaning a 
log-normal distribution in this log-log plot, for large 
language sizes. 

A simple scaling law, seen in Fig.4, predicts the behaviour of 
the number $L$ of languages as a function of the maximum possible
number $M$ of languages and the final population $N_\infty
\simeq K/2$:

$$ L/M = f(M/N_\infty) \quad .$$
The scaling function $f(z)$ equals unity for small $z$ and 
decays as $1/z$ for large $z$. This means that for a population
much larger than the possible number of languages, each language
possibility is realized, while in the opposite limit each small
group of individuals speaks its own language. Therefore we expect
this simple scaling law to be valid also for longer bit-strings
than the 8 and 16 bits simulated here. (32 bits allow for 4096 Mega
languages, requiring too much computer memory in our program;
30 bits still worked.)

We also modified the model to take into account the influence of a ``superior''
language on another, like the many words of French origin in the German 
language. With some probability $q$, at the moment of a mutation the new
value of a bit is not the opposite of the old value  (as done above) but is
the value of the corresponding bit in the superior language. We define
as superior language the bit-string having one everywhere except for a zero
in the left-most position, i.e. 127 for 8 and 32767 for 16 bits. The larger
$q$ is (in the smooth regime of large $p = 0.48$ per individual),
the higher is the fraction of samples ending with the superior language
as the largest one. About half of the samples have the superior language
as the numerically strongest one if $q \simeq 0.02$ for 8 and 0.2 for 16
bits. If for 16 bits we take 127 instead of 32767 as the superior language,
the results do not change much. (These probabilities hold for 10 million people
and are appreciably larger, 0.05 and 0.34, for one million.)

\section{Discussion}

Our model is more microscopic than the previous ones known to us
\cite{strogatz,finland} in that individuals are born, give 
birth, and die, instead of being lumped together into one
differential equation. It also is more realistic since we allow 
for numerous languages instead of only two. For the latter 
choice, we would have to reduce our bit-string to a single bit,
with $M = 2$ and thus $M/N \ll 1$, corresponding to the left 
part of Fig.4. There we observe $L = M$, that means both 
languages survive. In \cite{strogatz} only one language 
survived since one was assumed to be superior compared to the 
other. We, on the other hand, regarded all languages as 
intrinsically equally fit, except for the last paragraph.

\bigskip

We thank P.M.C. de Oliveira for suggesting to simulate languages, and the
Julich supercomputer center for JUMP time.


\begin{thebibliography}{99}
\bibitem{science} "Evolution of Language", special section in Science 303, 
1315-1335 (2004).
\bibitem{sutherland} W.J. Sutherland, Nature 423, 276 (2003).
\bibitem{eigen} M. Eigen, Naturwissenschaften 58, 465 (1971).
\bibitem{pmco} P.M.C. de Oliveira et al., Phys. Rev. E 70, 051910 (2004).
\bibitem{book} S. Moss de Oliveira, P.M.C. de Oliveira and D. Stauffer,
{\it Evolution, Money, War and Computers}, Teubner, Leipzig and Stuttgart 1999.
\bibitem{strogatz}  D.M. Abrams and S.H. Strogatz, Nature 424, 900 (2003); 
for more than two languages see
M.A. Nowak, N.L. Komarova and P. Niyogi, Nature 417, 611 (2002).
\bibitem{finland} M. Patriarca and T. Leppanen, Physica A 338, 296 (2004).

\end{thebibliography}
\end{document}